\documentclass[conference]{IEEEtran}
\IEEEoverridecommandlockouts

\usepackage{cite}
\usepackage{float}
\usepackage{amsmath,amssymb,amsfonts}
\usepackage{algorithmic}
\usepackage{multirow}
\usepackage{graphicx}
\usepackage{enumitem}
\usepackage{hyperref}
\usepackage{textcomp}
\usepackage{tcolorbox}
\tcbuselibrary{breakable}
\usepackage{wasysym}    
\usepackage{makecell}   
\usepackage{booktabs}   
\usepackage{xcolor}
\usepackage{url}
\def\BibTeX{{\rm B\kern-.05em{\sc i\kern-.025em b}\kern-.08em
    T\kern-.1667em\lower.7ex\hbox{E}\kern-.125emX}}

\newboolean{showcomments}
\setboolean{showcomments}{true}         
\ifthenelse{\boolean{showcomments}}
  {\newcommand{\nb}[2]{
  \fbox{\bfseries\sffamily\scriptsize#1}
     {\sf\small$\blacktriangleright$\textit{\textcolor{red}{#2}}$\blacktriangleleft$}
   }
  }
  {\newcommand{\nb}[2]{}
   
  }

\makeatletter
\newcommand{\linebreakand}{%
  \end{@IEEEauthorhalign}
  \hfill\mbox{}\par
  \mbox{}\hfill\begin{@IEEEauthorhalign}
}
\makeatother

\begin{document}

\title{Green Code or Costly Code? Evaluating the Energy Costs of Iterative LLM-Based Code Optimizations \\
\thanks{Accepted at the 49th IEEE/ACM International Conference on Software Engineering (ICSE 2027).}
}

\author{\IEEEauthorblockN{1\textsuperscript{st} Sadra Emami}
\IEEEauthorblockA{
\textit{University of Augsburg}\\
Augsburg, Germany \\
sadra.emami@uni-a.de}
\and
\IEEEauthorblockN{2\textsuperscript{nd} Benjamin Weigell}
\IEEEauthorblockA{
\textit{University of Augsburg}\\
Augsburg, Germany \\
benjamin.weigell@uni-a.de}
\and
\IEEEauthorblockN{3\textsuperscript{rd} Mohammad Reza Mousavi}
\IEEEauthorblockA{
\textit{King’s College London}\\
London, United Kingdom \\
mohammad.mousavi@kcl.ac.uk}
\linebreakand
\IEEEauthorblockN{4\textsuperscript{th} Gunel Jahangirova}
\IEEEauthorblockA{
\textit{King’s College London}\\
London, United Kingdom \\
gunel.jahangirova@kcl.ac.uk}
\and
\IEEEauthorblockN{5\textsuperscript{th} Hugo Araujo}
\IEEEauthorblockA{
\textit{King’s College London}\\
London, United Kingdom \\
hugo.araujo@kcl.ac.uk}
\and
\IEEEauthorblockN{6\textsuperscript{th} Bernhard Bauer}
\IEEEauthorblockA{
\textit{University of Augsburg}\\
Augsburg, Germany \\
bernhard.bauer@uni-a.de}
}

\maketitle

\begin{abstract}
\label{sec:abstract}
Large Language Models (LLMs) are increasingly used to optimize code for
performance. This practice often assumes that improvements in execution speed naturally translate into overall lower energy consumption. We challenge this assumption empirically by measuring whether performance-optimized code produced by LLM self-refinement pipelines
is also energy-efficient relative to canonical human reference solutions, and whether the energy cost of the optimization process can be offset by subsequent execution-time savings.
We evaluate 11 open-source models across four decoding configurations
on all 1,000 problems of EffiBench, a state-of-the-art performance benchmark, while collecting per-process, hardware-level energy measurements throughout. Inference energy
spans 0.47--6.50\,MJ across model-variant pairs, which is roughly equivalent
to the energy of 150--2100 GPT-4.1 queries. Limiting the maximum number of output tokens has the largest impact, reducing energy use by up to 72\% compared to the default configuration. On the execution side,
38 of 44 configurations produce code that consumes \emph{more} energy per
problem than the canonical human reference, with overheads ranging from
+6\% to +638\%. The six exceptions achieve only
marginal gains, with a median break-even point of about $2 \times 10^5$
executions per problem. Mixture-of-Experts
architectures consume substantially less inference energy than
equivalently-sized dense models without sacrificing correctness, making
architectural choice a more consequential sustainability lever than decoding
parameters. Performance optimization and energy efficiency are not always aligned
goals: LLM self-refinement consistently produces faster-seeking but
energy-heavier code relative to what a human engineer would write. All artifacts are available at 
\url{https://doi.org/10.5281/zenodo.23598169}.
\end{abstract}


\section{Introduction}
\label{sec:intro}

Recent studies indicate that developers routinely integrate Large Language
Models (LLMs) into their daily coding workflows~\cite{fan2023large,hou2024large}. Beyond code generation, LLMs are used to iteratively refine code to reduce
execution time and memory usage, tasks that previously required expert manual
effort~\cite{huang2024effilearner}. This growing practice implicitly assumes that performance improvements also translate into environmental benefits: that faster code is greener code. However, each optimization interaction carries two distinct energy costs: the \emph{inference energy} consumed when the model produces code~\cite{luccioni2024power,samsi2023words}, and the \emph{execution energy} consumed when that code is later run~\cite{pereira2017energy}. Both costs are largely invisible to the developer, yet together determine whether an optimization yields a net energy benefit. However, it remains unclear whether the execution-time savings can offset the energy required to generate the optimization, or whether LLM-optimized code is more energy-efficient than code written by human developers~\cite{islam2025evaluating,solovyeva2025ai}.

Coignion
et al.~\cite{coignion2025faster} introduce the Break-Even Point (BEP), defined
as the number of executions at which cumulative execution savings offset
generation cost, and find that this point ranges from under 100 to over 200,000
executions depending on the large language model and the method that is being optimized. However, their analysis measures savings relative to each model's own unoptimized output rather than a canonical human-written solution. This distinction is important because a model can greatly improve upon its own baseline while still producing code that is less efficient than a competent human solution. As a result, it remains unclear whether LLM-optimized code is actually competitive with what a competent human engineer would produce.

We investigate this question by evaluating the energy consumption of LLM-based code optimization end-to-end. We use EffiLearner~\cite{huang2024effilearner}, a state-of-the-art iterative self-refinement pipeline for code performance optimization, to generate optimized solutions. We evaluate these solutions against the canonical human-written reference implementations in EffiBench~\cite{huang2024effibench}, a benchmark of 1,000 LeetCode problems paired with top-rated human solutions. We evaluate only open-source models deployed on-premise. Such deployments are increasingly adopted in privacy-sensitive environments, while also allowing us to instrument the underlying hardware and obtain precise, hardware-level measurements of inference energy rather than relying on estimates or provider-reported statistics.

Addressing the practical value of LLM-based optimization requires examining three different aspects. First, the optimization process itself incurs an energy cost that must be quantified. Second, this cost may vary substantially across models and decoding strategies, potentially changing the trade-off between optimization quality and energy consumption. Finally, the generation investment is worthwhile only if the resulting code consumes less execution energy than a strong human-written baseline, allowing the initial energy cost to be recovered over repeated executions. These considerations motivate the following research questions:

\begin{itemize}
  \item \textbf{RQ1.} What is the energy cost of generating performance-optimized
    code using LLMs?

  \item \textbf{RQ2.} How does the energy cost of generating optimized code vary
    across open-source LLMs and decoding configurations?
    \begin{itemize}
      \item \textbf{RQ2a.} How do decoding parameters (temperature, context window,
        maximum output tokens) affect energy consumption across models?
      \item \textbf{RQ2b.} Which model-configuration pairs offer the most favorable
        trade-off between inference energy and optimization quality?
    \end{itemize}
  \item \textbf{RQ3.} Does performance-optimized code consume less execution
    energy than canonical human reference solutions, and if so, at what number
    of executions does the generation investment break even?
\end{itemize}

\noindent\textbf{Contributions.} 

Our main contributions are as follows:

\begin{enumerate}[leftmargin=*]

\item We present the first end-to-end energy evaluation of iterative LLM-based
code optimization against canonical human-written reference solutions.

\item We conduct a large-scale empirical study of the energy cost of LLM-based
code optimization across 11 open-source models and multiple decoding
configurations, providing a systematic characterization of the factors that
influence inference energy and optimization cost.

\item We release all experimental artifacts, including
repaired EffiBench benchmark, to
support reproducibility and future research on energy-efficient LLM-based code
optimization.

\end{enumerate}

\noindent\textbf{Findings in brief.} Our results show that inference energy ranges from
0.47 to 6.50\,MJ across model--configuration pairs, equivalent to roughly
150--2100 GPT-4.1 queries~\cite{jegham2025hungry}. Among the evaluated decoding parameters,
limiting the maximum output length has by far the greatest impact, reducing inference energy by 5--72\% relative to the default
configuration.

On the execution side, the optimized code is almost never more
energy-efficient than the canonical human reference: 38 of the 44 evaluated
configurations consume between 6\% and 638\% more execution energy per problem.
Only six configurations outperform the human baseline, requiring a
median of about $2.1 \times 10^{5}$ executions per problem (range
$1.0 \times 10^{5}$ to $8.4 \times 10^{5}$) to recover the energy invested in
optimization.

The rest of this paper is organized as follows. \textbf{Section~\ref{sec:related}} surveys related work. \textbf{Section~\ref{sec:methods}}
describes the dataset, pipeline, models, hardware, and instrumentation.
\textbf{Section~\ref{sec:results}} reports results for each RQ and the carbon-footprint
analysis. \textbf{Section~\ref{sec:discussion}} discusses the implications of our findings.
\textbf{Section~\ref{sec:limitations}} outlines the limitations of our study and directions for future work.
\textbf{Section~\ref{sec:conclusion}} draws the main conclusions of the study.
\section{Related Work}
\label{sec:related}

The energy footprint of LLMs has been studied along two axes: the cost of
\emph{operating} the model, and the cost of \emph{running the code} the model
produces. Our study sits at the intersection of both.

\subsection{Inference-Pipeline Energy}
\label{sec:rw-inference}

\begin{table*}[!t]
\centering
\caption{Coverage of the dimensions relevant to our three RQs across prior work, grouped by capability profile.\ \Circle\ = not addressed, \LEFTcircle\ = partial, \CIRCLE\ = addressed.}
\label{tab:related}
\footnotesize
\setlength{\tabcolsep}{4pt}
\renewcommand{\arraystretch}{1.3}
\begin{tabular}{l c c c c c c c c}
\toprule
Group & Task
 & \makecell{Inference\\Energy}
 & \makecell{Runtime\\Energy}
 & \makecell{Standardized\\Benchmark}
 & \makecell{Parameter\\Sweep}
 & \makecell{Break-Even\\Analysis}
 & \makecell{Iterative\\Pipeline}
 & \makecell{Canonical\\Reference} \\
\midrule
\makecell[l]{Inference-energy studies \\
\cite{luccioni2023estimating,samsi2023words,poddar2025towards,adamska2026greenpromptingcharacterizingpromptdriven}}
& General LLM & \CIRCLE & \Circle      & \Circle      & \LEFTcircle & \Circle     & \Circle  & \Circle \\
\makecell[l]{DL framework energy \\
\cite{georgiou2022green}}
& General DL & \CIRCLE & \Circle      & \CIRCLE      & \Circle     & \Circle     & \Circle  & \Circle \\
\makecell[l]{Inference energy for code gen. \\
\cite{rubei2025prompt,ashraf2025toward,ilager2025green,shi2024greening}}
& Code Generation   & \CIRCLE & \Circle      & \LEFTcircle  & \Circle     & \Circle     & \Circle  & \Circle \\
\makecell[l]{Generated-code energy \\
\cite{huang2024effibench,islam2025evaluating,tuttle2024can,apsan2025generating}}
& Code Generation   & \Circle & \CIRCLE      & \CIRCLE      & \Circle     & \Circle     & \Circle  & \CIRCLE \\
\makecell[l]{Code-optimisation pipelines \\
\cite{huang2024effilearner,dearing2025leveraging,rani2025can}}
& Code Optimization   & \Circle & \CIRCLE      & \LEFTcircle  & \Circle     & \Circle     & \CIRCLE  & \LEFTcircle \\
\makecell[l]{Break-even analysis \\
\cite{coignion2025faster}}
& Code Optimization   & \CIRCLE & \CIRCLE      & \CIRCLE      & \Circle     & \CIRCLE     & \LEFTcircle & \Circle \\
\midrule
\textbf{Our work}
& Code Optimization   & \CIRCLE & \CIRCLE      & \CIRCLE      & \CIRCLE     & \CIRCLE     & \CIRCLE  & \CIRCLE \\
\bottomrule
\end{tabular}
\end{table*}

Luccioni et al.~\cite{luccioni2023estimating} reported that pre-training BLOOM
consumed 433{,}196~kWh and 24.69~tonnes of CO\textsubscript{2}eq. The
systematic mapping by Verdecchia et al.~\cite{verdecchia2023systematic}
confirmed that the Green AI literature is heavily skewed toward training, with
inconsistent measurement protocols hampering cross-study comparison. Budennyy
et al.~\cite{budennyy2022eco2ai} released \texttt{eco2AI}, exposing a 26-fold
range in grid carbon intensity that makes \emph{where} a model runs nearly as
consequential as how efficiently it runs. These studies establish the
measurement landscape our work builds on, but focus exclusively on training. The inference cost of iterative optimization pipelines remains unexamined.

On the inference side, Samsi et al.~\cite{samsi2023words} provide the most
detailed power profile of LLaMA inference across deployments, observing system
draws between 300~W and 1~kW, while a study~\cite{de2023growing} estimates
that at ChatGPT-scale traffic, inference matches the energy cost of training
in just about 2~days. LASSI-EE~\cite{dearing2025leveraging} is the closest
prior pipeline to ours in spirit: an iterative, self-correcting LLM loop with
runtime power-profile feedback applied to HPC code, reporting 36\% and 34\%
energy reductions on MI100 and A100 respectively. Like the training-focused
studies above, however, it does not connect inference and execution energy
through a break-even analysis, and targets bespoke HPC kernels rather than a
standardised benchmark like EffiBench. Crucially, no prior work jointly
measures inference and execution energy across a controlled decoding parameter
sweep on a standardised benchmark, which is the gap this paper fills.

\subsection{Effect of Model Configuration on Inference Energy}
\label{sec:rw-config}

Several recent studies have identified output token length as the dominant
inference energy driver. Poddar et al.~\cite{poddar2025towards} report a
Pearson correlation of $r=0.952$ between output tokens and energy, against
$r=0.697$ for input tokens. Adamska et
al.~\cite{adamska2026greenpromptingcharacterizingpromptdriven} independently
measure $r=0.9$ on a separate corpus and find that prompt phrasing can influence response length. Both findings emerge as observational
correlations across natural output variation rather than from controlled
ablations of the output-token budget as an independent variable. To our 
knowledge, no prior study provides an isolated, controlled measurement 
of \emph{sampling temperature}, \emph{context window size}, or \emph{maximum 
output tokens} as independent inference-energy levers while holding the 
remaining factors constant. Furthermore, both studies focus on general NLP
tasks such as classification, question answering, and summarization; neither
examines code generation or optimization pipelines, where output structure and
length distributions differ substantially from natural language responses. Controlled comparisons do exist for the software stack. Georgiou et
al.~\cite{georgiou2022green} measure training and inference energy of six deep
learning models under matched configurations in PyTorch and TensorFlow, and
report statistically significant differences between the frameworks in 94\% of
the energy and runtime comparisons, with comparable accuracy and no single
framework superior across all tasks. The factor they vary is the framework,
however, and the models they study are not applied to code generation.

\subsection{Runtime Energy of Generated Code and the Break-Even Point}
\label{sec:rw-codeenergy}

Empirical consensus establishes that LLM-generated code is systematically 
less efficient than canonical human solutions. EffiBench~\cite{huang2024effibench},
which evaluated 42 models, found that even the best open-source model
(StarCoder2-15B) required 2.59$\times$ the execution time and 1.71$\times$
the peak memory of human solutions. Islam et al.~\cite{islam2025evaluating}
extend this picture on 878 EffiBench problems, finding that the best model
(DeepSeek-v3) remains 1.17$\times$ less energy-efficient than humans, with
gaps reaching 449$\times$ for dynamic programming and bit-manipulation tasks.
Convergent findings come from Tuttle et al.~\cite{tuttle2024can}, who report
LLMs outperforming human solutions in only 5.7\% of 192 LeetCode responses.
Apsan et al.~\cite{apsan2025generating}, who show the human--LLM gap persists
across server, PC, and Raspberry Pi platforms. Solovyeva et
al.~\cite{solovyeva2025ai}, who find parity only in narrow regimes such as
String and Tree tasks; and Rani et al.~\cite{rani2025can}, who document that
LLM-suggested optimizations on Matlab often unintentionally increase memory
and energy usage. Whether iterative self-refinement can close this gap, and at what inference 
cost, motivates the methodological thread we build on.

The methodological precursor to our pipeline is
EffiLearner~\cite{huang2024effilearner}, which iteratively refines
LLM-generated code by feeding execution-overhead profiles back to the model as
structured prompts, demonstrating consistent reductions in Normalized Execution Time (NET) and Normalized Total Memory Usage (NTMU) across
22 models, but without measuring inference energy and therefore unable to
answer whether the optimization investment is recovered at execution time.
Coignion et al.~\cite{coignion2025faster} first formalised this question via
the BEP ($BEP = E_{\text{optim}} / (e_{\text{orig}} - e_{\text{opt}})$),
finding results spanning four orders of magnitude: DeepSeek-R1-14B with EoH
yields 49\% per-execution savings but requires 220{,}324 executions to break
even, while pathologically inefficient code can drop from 318~J to 2.6~J and
break even within 100 runs. Critically, their BEP measures savings relative to
each model's own unoptimized output rather than a canonical human reference,
leaving open the question of whether LLM-optimized
code is competitive with what a human engineer would write.

\subsection{Gap Analysis}
\label{sec:rw-positioning}

Most of the above studies address either (i) the inference energy of LLMs in
general NLP settings, (ii) the runtime efficiency of LLM-generated code
relative to human references, or (iii) the break-even economics of LLM-based
optimization relative to each model's own unoptimized output. No prior study
jointly measures inference energy across a controlled decoding parameter sweep,
evaluates the resulting code against canonical human solutions, and applies a
break-even analysis within a multi-epoch self-refinement pipeline on a
standardised benchmark. Our study is designed to fill this intersection.

Table~\ref{tab:related} summarizes how our study differs from the closest
prior work along the dimensions most relevant to our three RQs.

\section{Methods}
\label{sec:methods}

In this section, we first describe the benchmark dataset and the self-refinement framework. We then present the evaluated models and decoding configurations, followed by the hardware platform and energy-measurement instrumentation. Finally, we explain how this experimental setup is used to answer each research question.

\subsection{Dataset}
\label{sec:dataset}

We evaluate on EffiBench~\cite{huang2024effibench}, a benchmark of 1,000 efficiency-critical LeetCode problems spanning 171 \emph{Easy}, 589 \emph{Medium}, and 240 \emph{Hard} problems. Each entry pairs a problem description with an executable canonical solution drawn from top-starred LeetCode Discussion Forum submissions, a test-case generator, and 100 test cases. The canonical solutions serve as the human reference baseline for our RQ3
break-even analysis. Because they are top-starred submissions, they represent
highly efficient human implementations rather than average developer code, and
we discuss the implications in Section~\ref{sec:limitations}.

EffiBench is particularly well suited to our study because its algorithmically challenging problems exhibit substantial variation in execution time, making differences in execution energy measurable. 

The EffiBench test cases were generated by GPT-3.5 Turbo, and during our
validation run we found that 122 of the 1{,}000 problems contained defects
preventing correct execution, all originating in the generated test cases
rather than the solutions themselves.
The most common defects were ordering-sensitive
assertions on unordered outputs, missing helper utilities for tree and
linked-list construction, object-address serialization embedding raw in-memory
representations, wrong precomputed expected values, method-name mismatches,
redundant class redefinitions shadowing the actual implementation, and empty
test cases caused by generator logic that emitted no output. For each defect
type we applied a targeted repair, recomputing expected values from the
canonical solution, injecting missing helpers, rewriting object references as
constructor calls, removing shadowing definitions, and converting
ordering-sensitive assertions to order-independent equivalents. After repair,
all 1{,}000 canonical solutions pass their respective test cases.

\subsection{Self-Refinement Framework}
\label{sec:effilearner}
We adopt EffiLearner~\cite{huang2024effilearner} as our self-refinement pipeline because it directly targets execution-time and memory efficiency through line-level profiling feedback, and has been shown to produce significantly more efficient code across a wide range of model architectures, achieving state-of-the-art reductions in execution time and memory usage compared to alternative self-refinement approaches. EffiLearner structures code optimization as an iterative refinement loop comprising three stages: code generation, overhead profiling, and code refinement. Given a task description, the LLM first produces an initial code candidate.

At each
iteration, the current candidate is executed locally against the set of open
test cases provided by the benchmark, and two line-level overhead profiles are
collected: an execution-time profile via \texttt{line\_profiler} and a
memory-usage profile via \texttt{memory\_profiler}. Each profiling measurement is repeated 60 times per script, and the
reported Execution Time (ET), Max Memory Usage (MU), and Total Memory Usage (TMU) values are the averages over those 60 runs.
To minimize the influence of thermal throttling and background load, a
cooldown period of 15 seconds is enforced between consecutive runs. A run is only admitted once CPU load falls below 50\% and the CPU
package temperature is within the 50--60\,\textdegree C range. Both profiles aggregate
measurements across all open test cases and report, per line, the cumulative
time and memory footprint. The current candidate, its overhead profiles, the
task description, and the open test cases are then concatenated into a
refinement prompt and passed back to the LLM, which emits a revised candidate.
The revised candidate is validated for functional correctness against the open
test cases only; if it passes, it replaces the current candidate as the input
to the next iteration. The loop terminates after a fixed number of iterations.
Final evaluation is conducted on a held-out set of private test cases disjoint
from the open test cases, following the protocol of
EffiBench~\cite{huang2024effibench}.

\subsection{Models and Parameter Configurations}
\label{sec:sweep}

Models were selected by querying the Arena Leaderboard on Hugging
Face\footnote{\url{https://huggingface.co/spaces/lmarena-ai/arena-leaderboard},
accessed March 2026} and sorting by coding performance. We restricted our
selection to open-source models deployed entirely on-premise, as local
inference enables controlled energy measurement on a fixed, instrumented
hardware environment and reflects the growing adoption of local deployment in
privacy-sensitive settings. We selected representative size variants from the
top-performing families, ensuring architectural diversity across both dense and Mixture-of-Experts (MoE) architectures to enable a controlled comparison of their energy and quality
characteristics. The resulting set of 11 models spans four
families~\cite{yang2025qwen3,glm2024chatglm,liu2026ministral,wang2025nemotron,blakeman2025nemotron},
different size scales, and two architectural classes, as listed in
Table~\ref{tab:models}.

\begin{table}[htbp]
\caption{Models evaluated in this work.}
\label{tab:models}
\centering
\begin{tabular}{llcc}
\hline
\textbf{Family} & \textbf{Model} & \textbf{Parameters} & \textbf{Architecture} \\
\hline
\multirow{4}{*}{Qwen3.5}   & Qwen3.5-4B       & 4B  & Dense \\
                            & Qwen3.5-9B       & 9B  & Dense \\
                            & Qwen3.5-27B      & 27B & Dense \\
                            & Qwen3.5-35B-A3B  & 35B & MoE   \\
\hline
\multirow{2}{*}{GLM}       & GLM-4.7-Flash    & 31B & Dense \\
                            & GLM-4.6V-Flash   & 10B & Dense \\
\hline
\multirow{3}{*}{Ministral} & Ministral-3B     & 3B  & Dense \\
                            & Ministral-8B     & 8B  & Dense \\
                            & Ministral-14B    & 14B & Dense \\
\hline
\multirow{2}{*}{Nemotron}  & Nemotron-Cascade & 30B & MoE \\
                            & Nemotron-Nano    & 30B & MoE \\
\hline
\end{tabular}
\end{table}

For each model, we run four configurations that each change at most one
parameter, so that energy differences can be attributed to a single cause.
We vary only deployment levers that practitioners can adjust without
retraining and that act on energy through the number of generated tokens or
the memory provisioned for generation. The \emph{default} configuration uses
temperature 0 (greedy decoding) and a maximum output length of 2{,}048
tokens. The \emph{sampling} variant raises the temperature to $0.5$. In a
pilot experiment, temperatures above $0.5$ increased verbosity and inference
energy without improving code efficiency, so we did not sweep higher values.
The \emph{reduced-context} variant caps the context window at 64k tokens,
since all evaluated models support windows three to four times larger by
default. The \emph{reduced-output} variant halves the maximum output length
to 1{,}024 tokens to test whether a shorter generation budget lowers energy
without hurting optimization quality. A full factorial sweep was timely infeasible.

\subsection{Experimental Setup}
\label{sec:hardware}

All experiments were conducted on a server equipped with two Intel Xeon
Gold 6438Y+ processors with 128 logical cores in total, 1\,TB of system memory,
and two NVIDIA L40 GPUs with 46\,GB GDDR6 VRAM each. Simultaneous
Multithreading (SMT) was enabled and the system exposed two Non-Uniform Memory
Access (NUMA) nodes. The server ran Ubuntu 24.04 with Linux kernel 6.8.0-138,
and models were served locally via vLLM~\cite{kwon2023efficient}. Processes
were pinned to physical cores, while the frequency governor, turbo boost,
C-states, and NUMA policy remained at their operating system defaults to
reflect standard deployments. The full experimental pipeline required
approximately 30 days of continuous execution.

\subsection{Energy Measurement}
\label{sec:energy}

We instrument all experiments with Metrion~\cite{weigell2025metrion}, an
energy measurement tool that attributes energy consumption to processes at the
thread level. Most existing tools neglect the combined effects of frequency
scaling, simultaneous multithreading, and non-uniform memory access in
multi-socket architectures. Utilization-based tools like
CodeCarbon~\cite{lacoste2019quantifying} estimate process-level energy from
coarse-grained CPU time and RAPL values without accounting for these
hardware-level effects, while profiling utilities like \texttt{perf} expose
RAPL metrics only at the package level. Metrion overcomes these limitations by
combining Intel RAPL, nvidia-smi, and Hardware Performance Counters (HWPCs) to
attribute CPU, DRAM, and GPU energy at the thread level, with HWPCs resolving
per-thread footprints within each 1-second sampling window. Metrion does not
require a process to span an entire sampling interval. It uses eBPF and HWPCs
to trace when each process is scheduled on or off the CPU and to count its
unhalted cycles, and it allocates the RAPL package energy, sampled at 1\,Hz, to
active processes in proportion to their unhalted cycles in each interval. This
attributes energy correctly even to short-lived processes. Frequency scaling
and hardware topology are handled at attribution time. Metrion weights active
cycles by the APERF/MPERF ratio, which reflects the actual frequency under
Dynamic Voltage and Frequency Scaling (DVFS) and Turbo Boost, scales
overlapping SMT threads by a calibrated contention factor of 1.15, and
attributes energy per socket. Energy is aggregated over all descendant threads
of each solution process, yielding a per-solution breakdown into CPU, DRAM, and
GPU, each decomposed into active and idle components.

\subsubsection{Measurement Protocol}
\label{sec:protocol}

Inference energy and execution energy are measured under different
repetition budgets, reflecting the difference in their computational
cost.

\paragraph{Inference energy.}
Each \emph{(model, configuration)} pair is executed four times in full,
covering all 1{,}000 problems per run. The reported $E_{\text{total}}$
($= E_{\text{gen}} + E_{\text{exec}}$)
is the mean over these four runs. Running inference four times
provides sufficient stability for the aggregate energy signal while
keeping total experiment time feasible given the cost of a full
five-epoch optimization pass. GPU energy applies only to this inference phase,
whose runs last from hours to days, and is recorded continuously via NVML.

\paragraph{Execution energy.}
Each solution script is executed 60 times sequentially within one persistent
Python interpreter, and the reported $E_{\text{exec}}$ is the mean over those
60 runs. Keeping the interpreter alive amortizes startup and import overhead,
so the measurement reflects steady-state energy. Canonical and LLM-generated
solutions share an identical harness, so any residual overhead affects both
symmetrically. Individual executions average more than 0.1 seconds, so the 60
sequential executions of each solution give cumulative measurement windows well
over one second. The higher repetition count relative to inference is feasible
because individual script executions are cheaper. The same cooldown
protocol applied to resource profiling is used here: a 15-second
pause is enforced between consecutive executions, and a run is only
admitted once CPU load is below 50\% and the CPU package temperature
is within the 50--60\,\textdegree C range. These admission criteria prevent
cross-experiment interference and thermal throttling.

\subsection{Approaches to the Research Questions}
\label{sec:rqs}
The infrastructure described above produces two complementary energy
measurements at different granularities. For each
\emph{(model, configuration)} pair, we record the aggregate
\emph{generation energy} $E_{\text{gen}}$ expended by the LLM during
the entire generation and optimization run across all 1{,}000 problems. For each \emph{(model, configuration, problem)}
triple, it additionally records the per-process \emph{execution
energy} $E_{\text{exec}}$ consumed when running the resulting
solution script. We use
these measurements to answer each research question as follows.

\subsubsection{RQ1 -- Energy Cost of Optimization}
To quantify the energy cost of generating performance-optimized code,
we report $E_{\text{gen}}$ at the granularity at which it is measured:
one aggregate value per \emph{(model, configuration)} pair, broken
down by hardware component (CPU package, DRAM, GPU) and by dynamic
versus idle contributions, where the dynamic component captures power
draw above the static baseline and the idle component captures the
baseline itself. $E_{\text{gen}}$ includes the full generation cost
across all five optimization epochs, including the initial code
generation. We report an aggregate over all 1{,}000 problems because vLLM batches
requests across problems and our research questions concern
model-level and configuration-level energy budgets.

\subsubsection{RQ2 -- Cross-Model and Parameter Variation}
RQ2 examines how generation energy varies with model choice and
decoding parameters. We evaluate this along two complementary axes.

\paragraph{Parameter sensitivity (RQ2a).} For every model in
Table~\ref{tab:models}, we compare the energy consumption of each model based on four parameter configurations
defined in Section~\ref{sec:sweep}. Holding the model fixed isolates the energy
impact of temperature, context size, and maximum output length.

\paragraph{Cross-model comparison (RQ2b).}
To identify configurations that simultaneously minimize generation cost and
improve the performance of the optimized code, we plot correctness efficiency
(pass@1/$E_{\text{total}}$, in \% MJ$^{-1}$) against the NET of the
last-epoch solutions for each of the four decoding configurations
(\texttt{default}, \texttt{t0.5}, \texttt{ctx65536}, and \texttt{mt1024}),
where $E_{\text{total}} = E_{\text{gen}} + E_{\text{exec}}$ is the sum of
generation and execution energy.

\subsubsection{RQ3 -- Break-Even of Generation vs.\ Execution Savings}
RQ3 asks whether the energy invested in generating optimized code is
recovered through reduced energy consumption at execution time. We
treat the canonical solution provided by EffiBench as the reference
baseline against which optimized solutions are compared. Each canonical
solution is a top-rated, human-written solution selected from
competitive programming submissions, representing a strong but
attainable efficiency ceiling. We aggregate at the (model,
configuration) level rather than per task because our goal is to
characterize the overall energy trade-off of deploying a given model
under a given decoding configuration, not to identify which individual
problems benefit from refinement.

For each \emph{(model, configuration)} pair, we compute the mean
execution energy of the optimized solutions
$\bar{E}_{\text{exec}}^{*}$ and the mean execution energy of the
canonical solutions $\bar{E}_{\text{exec}}^{\text{canon}}$ over the
benchmark, and define the per-run saving as
\[
    \Delta E = \bar{E}_{\text{exec}}^{\text{canon}}
             - \bar{E}_{\text{exec}}^{*}.
\]
The break-even point is the number of executions at which the
cumulative execution-time saving offsets the up-front generation
cost,
\[
N^{*} = \left\lceil \frac{E_{\mathrm{total}}}{n\,\Delta E} \right\rceil
\quad \text{when } \Delta E > 0,
\]
where $E_{\mathrm{total}}$ is the aggregate generation energy of the
optimization run and $n$ is the number of problems with a valid optimized
solution. $N^{*}$ is thus a single scalar per
\emph{(model, configuration)} pair, expressing how many executions
of a representative benchmark problem are required before the energy
invested in optimization is recouped relative to the canonical
baseline. When the mean optimized
solution is no more energy-efficient than the canonical baseline
($\Delta E \leq 0$), no break-even point exists and the optimization
constitutes a net energy loss in aggregate. We report this outcome
explicitly where it occurs.

\section{Results}
\label{sec:results}

This section presents results for three research questions\footnote{Due to a bug discovered in Metrion, the inference runs were repeated with the fixed version. Some energy consumption values changed as a result, but none of the conclusions did.}: the energy cost of LLM-based code optimization (Section~\ref{sec:rq1}), how that cost varies across models and decoding configurations (Section~\ref{sec:rq2}), and whether the generation investment breaks even against execution-time savings relative to canonical solutions (Section~\ref{sec:rq3}).

\subsection{RQ1: What is the energy cost of generating performance-optimized code using LLMs?}
\label{sec:rq1}

All 11 models first produced an initial solution through base generation, which was
then subjected to five epochs of iterative self-optimization. We evaluate functional
correctness via the pass rate on the private test cases, and resource efficiency via ET,
MU, and TMU, each reported in absolute terms and normalized against the profiling results
of the canonical solutions, resulting in NET, Normalized Max Memory Usage (NMU) and NTMU. Table~\ref{tab:rq1_resources} reports base generation values
(top row) and post-optimization values (bottom row) for all metrics, with the relative
change in parentheses. 

\paragraph{Execution Time.}
NET at base generation ranges from 0.97 (Ministral-14B) to 1.43
(Nemotron-Nano), with the GLM and Nemotron families consistently above 1.0.
After self-optimization, ET reductions of up to 29.0\% are achieved, with
Nemotron-Nano showing the largest relative improvement.
The Qwen family already starts below or near 1.0 at base generation and shows modest
further reductions of 1.5--1.7\%.

\paragraph{Memory Usage.}
Normalized Max Memory Usage (NMU) at base generation ranges from 1.0029 (GLM-4.6V-Flash) to 1.0411
(Nemotron-Nano). After self-optimization, some models reduce NMU by up to 2.1\%,
converging toward the canonical footprint. NTMU reductions are more pronounced,
reaching 36.6\% for Nemotron-Nano, and several models, including Ministral-3B,
Ministral-8B and GLM-4.7-Flash achieve NTMU values below 1.0 after
optimization.

\paragraph{Functional Correctness.}
The highest pass@1 is achieved by Qwen3.5-27B (81.0\%), followed by
Qwen3.5-35B-A3B (78.55\%) and Qwen3.5-9B (66.80\%). GLM models achieve
the lowest correctness, with GLM-4.7-Flash at 25.02\% and GLM-4.6V-Flash at
25.42\%.

\paragraph{Inference Energy.}
Table~\ref{tab:rq1_energy} reports the inference energy consumed
during the five-epoch optimization loop and shows that total inference energy varies considerably across models. Ministral-3B and Qwen3.5-4B are the most energy-efficient at
0.87\,MJ each. Qwen3.5-27B incurs the
highest cost at 6.43\,MJ, reflecting its larger parameter count. GLM-4.7-Flash
is the second most expensive at 3.43\,MJ despite achieving only 25.02\% pass@1,
making it the least cost-effective configuration overall. Notably, Qwen3.5-35B-A3B
consumes only 2.31\,MJ despite having more total parameters than Qwen3.5-27B. This is due to its MoE architecture,
which activates only a subset of parameters per token and therefore reduces
the effective compute per inference step. GPU energy dominates
across all models, accounting for 69--82\% of $E_{\mathrm{total}}$, with CPU
contributing 16--29\% and DRAM a small fraction (1.6--2.8\%).

\begin{table*}[!ht]
\caption{%
    Resource efficiency before (top row) and after (bottom row)
    self-optimization and pass@1, with relative change in parentheses.
    ET: Execution Time (s); MU: Max Memory Usage (MB); TMU: Total Memory Usage (MB$\cdot$s).
    NET, NMU and NTMU are ET, MU and TMU normalized by the canonical solution;
    values below 1.0 beat the canonical reference.
}
\label{tab:rq1_resources}
\centering
\renewcommand{\arraystretch}{0.9}
\setlength{\tabcolsep}{2pt}
\footnotesize
\begin{tabular}{lcccccccc}
\hline
\textbf{Model}
  & \textbf{ET (s)}
  & \textbf{NET}
  & \textbf{MU (MB)}
  & \textbf{NMU}
  & \textbf{TMU}
  & \textbf{NTMU}
  & \textbf{pass@1 (\%)} \\
\hline
\multirow{2}{*}{GLM-4.6V-Flash}
  & 0.133 & 1.052 & 23.098 & 1.003 & 2.215 & 1.053 & \multirow{2}{*}{25.4} \\
  & 0.131 {\small(1.4\%)} & 1.038 {\small(1.4\%)} & \textbf{23.091} {\small(0.0\%)} & \textbf{1.003} {\small(0.0\%)} & 2.189 {\small(1.2\%)} & 1.040 {\small(1.3\%)} & \\
\hline
\multirow{2}{*}{GLM-4.7-Flash}
  & 0.129 & 1.034 & 23.150 & 1.005 & 2.168 & 1.039 & \multirow{2}{*}{25.0} \\
  & 0.123 {\small(4.7\%)} & 0.984 {\small(4.8\%)} & 23.133 {\small(0.1\%)} & 1.005 {\small(0.1\%)} & 2.062 {\small(4.9\%)} & 0.989 {\small(4.8\%)} & \\
\hline
\multirow{2}{*}{Ministral-3B}
  & 0.126 & 0.994 & 23.439 & 1.018 & 2.170 & 1.024 & \multirow{2}{*}{32.7} \\
  & 0.120 {\small(4.8\%)} & \textbf{0.946} {\small(4.9\%)} & 23.248 {\small(0.8\%)} & 1.009 {\small(0.8\%)} & 2.026 {\small(6.7\%)} & \textbf{0.956} {\small(6.7\%)} & \\
\hline
\multirow{2}{*}{Ministral-8B}
  & 0.135 & 1.089 & 23.635 & 1.026 & 2.425 & 1.167 & \multirow{2}{*}{52.6} \\
  & \textbf{0.118} {\small(12.6\%)} & 0.949 {\small(12.9\%)} & 23.385 {\small(1.1\%)} & 1.015 {\small(1.1\%)} & 2.005 {\small(17.3\%)} & 0.966 {\small(17.2\%)} & \\
\hline
\multirow{2}{*}{Ministral-14B}
  & 0.121 & 0.971 & 23.303 & 1.011 & 2.031 & 0.980 & \multirow{2}{*}{55.3} \\
  & 0.119 {\small(1.7\%)} & 0.954 {\small(1.8\%)} & 23.231 {\small(0.3\%)} & 1.008 {\small(0.3\%)} & \textbf{1.996} {\small(1.7\%)} & 0.963 {\small(1.8\%)} & \\
\hline
\multirow{2}{*}{Nemotron-Cascade}
  & 0.142 & 1.143 & 23.743 & 1.030 & 2.529 & 1.213 & \multirow{2}{*}{43.2} \\
  & 0.130 {\small(8.4\%)} & 1.050 {\small(8.2\%)} & 23.480 {\small(1.1\%)} & 1.019 {\small(1.1\%)} & 2.234 {\small(11.7\%)} & 1.079 {\small(11.1\%)} & \\
\hline
\multirow{2}{*}{Nemotron-Nano}
  & 0.180 & 1.431 & 23.985 & 1.041 & 3.453 & 1.641 & \multirow{2}{*}{41.5} \\
  & 0.128 {\small(\textbf{29.0\%})} & 1.020 {\small(\textbf{28.7\%})} & 23.476 {\small(\textbf{2.1\%})} & 1.019 {\small(\textbf{2.1\%})} & 2.190 {\small(\textbf{36.6\%})} & 1.045 {\small(\textbf{36.4\%})} & \\
\hline
\multirow{2}{*}{Qwen3.5-4B}
  & 0.124 & 0.988 & 23.307 & 1.011 & 2.069 & 0.995 & \multirow{2}{*}{51.3} \\
  & 0.122 {\small(2.1\%)} & 0.972 {\small(1.6\%)} & 23.283 {\small(0.1\%)} & 1.010 {\small(0.1\%)} & 2.024 {\small(2.1\%)} & 0.979 {\small(1.6\%)} & \\
\hline
\multirow{2}{*}{Qwen3.5-9B}
  & 0.127 & 1.020 & 23.303 & 1.011 & 2.126 & 1.029 & \multirow{2}{*}{66.8} \\
  & 0.126 {\small(1.1\%)} & 1.003 {\small(1.7\%)} & 23.298 {\small(0.0\%)} & 1.011 {\small(0.0\%)} & 2.082 {\small(2.0\%)} & 1.012 {\small(1.7\%)} & \\
\hline
\multirow{2}{*}{Qwen3.5-27B}
  & 0.124 & 0.989 & 23.184 & 1.006 & 2.035 & 0.993 & \multirow{2}{*}{\textbf{81.0}} \\
  & 0.122 {\small(1.5\%)} & 0.974 {\small(1.6\%)} & 23.173 {\small(0.0\%)} & 1.005 {\small(0.0\%)} & 2.003 {\small(1.6\%)} & 0.977 {\small(1.6\%)} & \\
\hline
\multirow{2}{*}{Qwen3.5-35B-A3B}
  & 0.123 & 0.986 & 23.207 & 1.007 & 2.030 & 0.989 & \multirow{2}{*}{78.6} \\
  & 0.121 {\small(1.8\%)} & 0.971 {\small(1.5\%)} & 23.197 {\small(0.0\%)} & 1.006 {\small(0.0\%)} & 1.998 {\small(1.6\%)} & 0.974 {\small(1.5\%)} & \\
\hline
\end{tabular}
\end{table*}

\begin{table}[H]
\caption{%
    Inference energy consumption per model during self-optimization. Share is each component's percentage
    of $E_{\mathrm{total}}$.%
}
\label{tab:rq1_energy}
\centering
\scriptsize
\setlength{\tabcolsep}{3pt}
\begin{tabular}{lrrrrrrr}
\hline
 & \multicolumn{4}{c}{\textbf{Energy (MJ)}} & \multicolumn{3}{c}{\textbf{Share (\%)}} \\
\cline{2-5} \cline{6-8}
\textbf{Model}
  & $E_{\mathrm{total}}$ & $E_{\mathrm{CPU}}$ & $E_{\mathrm{DRAM}}$ & $E_{\mathrm{GPU}}$
  & \textbf{CPU} & \textbf{DRAM} & \textbf{GPU} \\
\hline
GLM-4.6V-Flash   & 1.050 & 0.268 & 0.026 & 0.755 & 25.6 & 2.5 & 71.9 \\
GLM-4.7-Flash    & 3.429 & 0.553 & 0.054 & 2.822 & \textbf{16.1} & \textbf{1.6} & 82.3 \\
Ministral-3B     & \textbf{0.871} & 0.248 & 0.025 & \textbf{0.599} & 28.5 & 2.8 & \textbf{68.7} \\
Ministral-8B     & 1.635 & 0.391 & 0.034 & 1.209 & 23.9 & 2.1 & 74.0 \\
Ministral-14B    & 2.313 & 0.539 & 0.042 & 1.732 & 23.3 & 1.8 & 74.9 \\
Nemotron-Cascade & 1.651 & 0.363 & 0.032 & 1.255 & 22.0 & 2.0 & 76.0 \\
Nemotron-Nano    & 1.721 & 0.363 & 0.031 & 1.327 & 21.1 & 1.8 & 77.1 \\
Qwen3.5-4B       & \textbf{0.871} & \textbf{0.230} & \textbf{0.022} & 0.620 & 26.4 & 2.5 & 71.1 \\
Qwen3.5-9B       & 1.595 & 0.362 & 0.033 & 1.200 & 22.7 & 2.1 & 75.2 \\
Qwen3.5-27B      & 6.426 & 1.194 & 0.104 & 5.128 & 18.6 & \textbf{1.6} & 79.8 \\
Qwen3.5-35B-A3B  & 2.314 & 0.451 & 0.040 & 1.822 & 19.5 & 1.7 & 78.8 \\
\hline
\end{tabular}
\end{table}

\begin{tcolorbox}[
  colback=gray!8,
  colframe=gray!40,
  boxrule=0.5pt,
  arc=3pt,
  left=6pt, right=6pt, top=5pt, bottom=5pt
]
\textbf{RQ1:}
Self-optimization consistently reduces ET, MU, and TMU across all models, with the Nemotron family showing the largest relative
improvements. Pass@1 varies substantially across
families, with Qwen models achieving the highest correctness. Inference energy is
strongly driven by model size, though MoE models consume considerably less energy
than equivalently-sized dense models. The GLM family combines high inference cost
with the lowest correctness, making it the least cost-effective choice overall.
\end{tcolorbox}


\subsection{RQ2: How does the energy cost of generating optimized code vary across open-source LLMs and decoding configurations?}
\label{sec:rq2}

\begin{figure*}[!ht]
\centering
\includegraphics[width=0.7\textwidth]{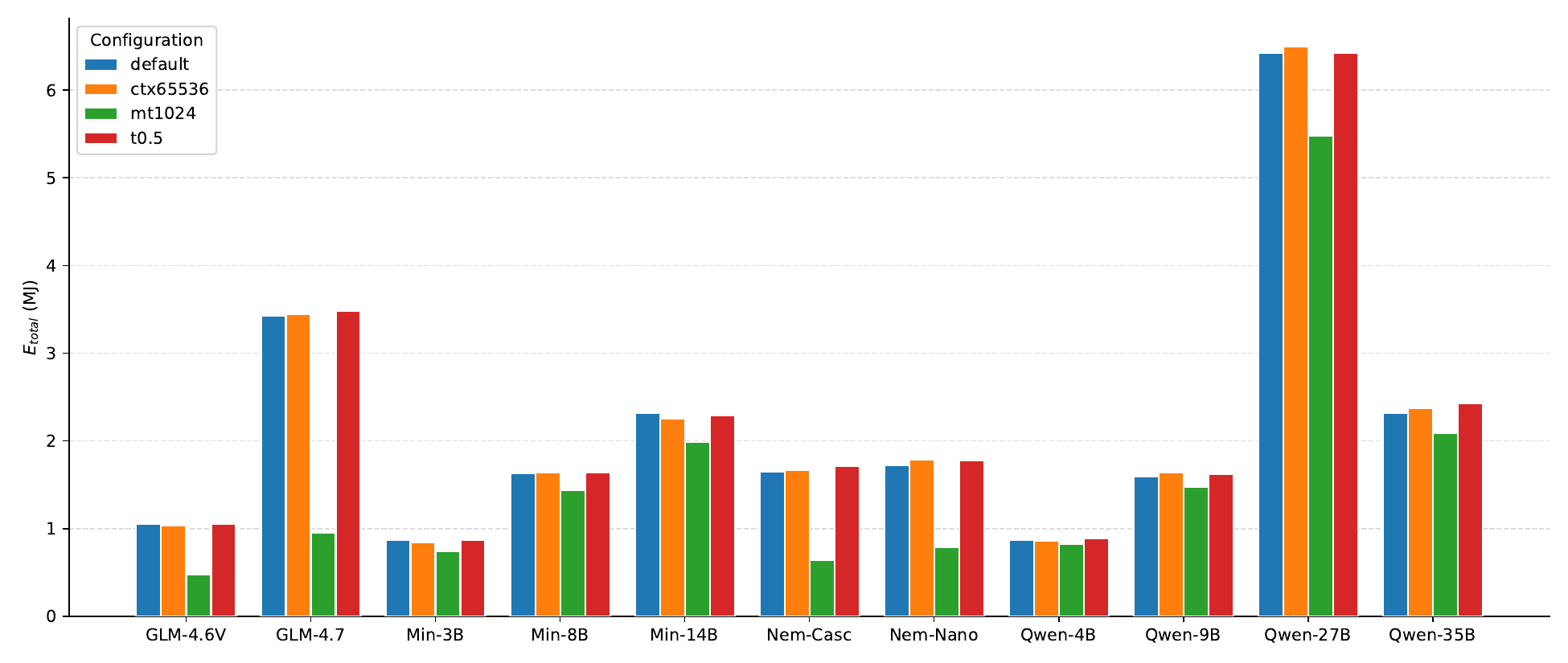}
\caption{%
    Total inference energy $E_{\mathrm{total}}$ (MJ) per model and decoding variant
    after five epochs of self-optimization. \textit{mt1024} is the lowest-energy
    variant for every model, and \textit{t0.5} is the highest for about half of them.
}
\label{fig:rq2_energy}
\end{figure*}

\begin{figure}[!ht]
\centering
\includegraphics[width=\columnwidth]{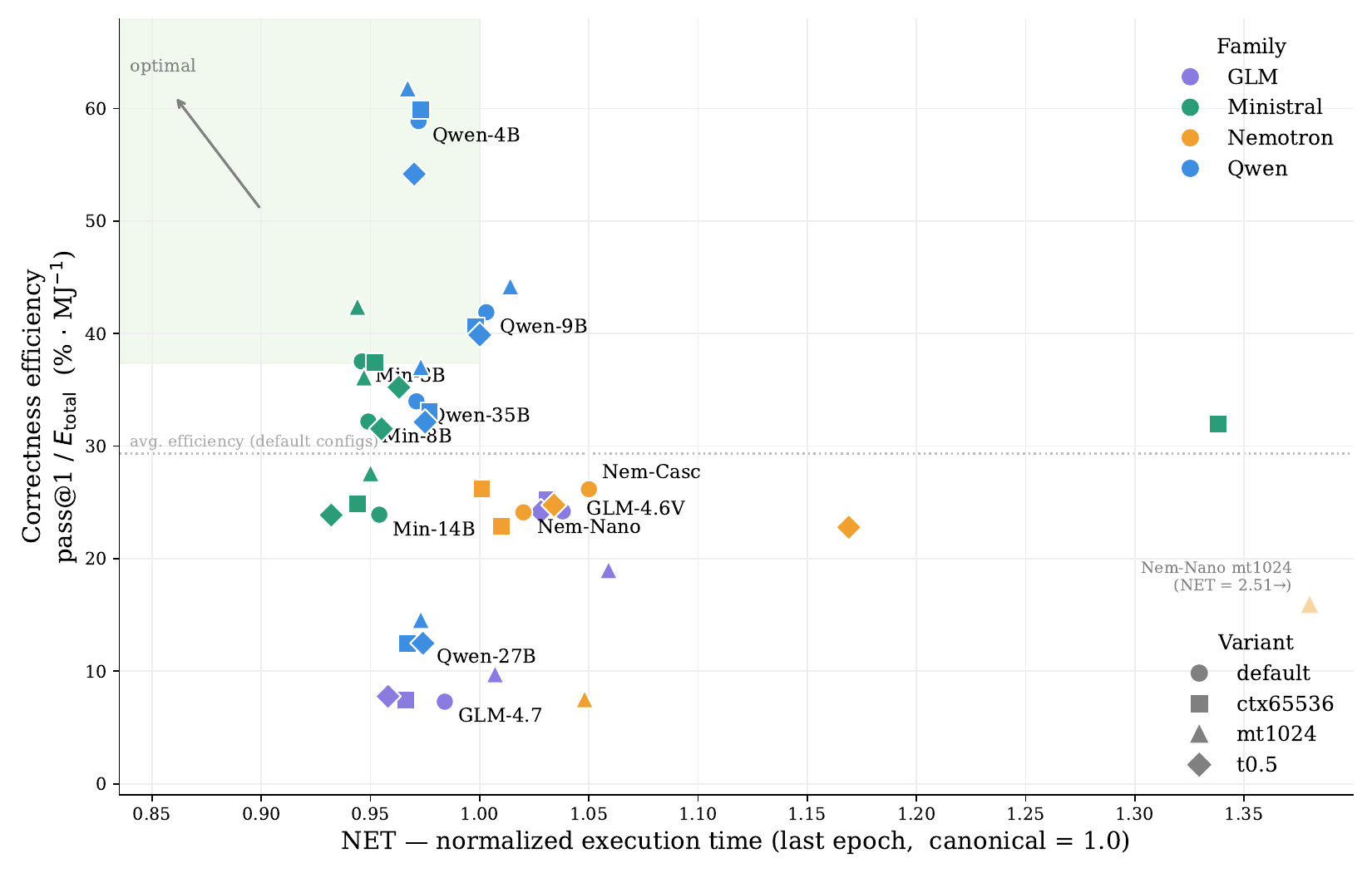}
\caption{%
    Correctness efficiency (pass@1\,/\,$E_{\mathrm{total}}$) vs.\ NET for all
    model-variant pairs. Shaded region: optimal zone. Dashed line: canonical
    baseline (NET\,=\,1.0). Nemotron-Nano \textit{mt1024} is clipped due to
    pass@1 collapse.
}
\label{fig:efficiency}
\end{figure}

Table~\ref{tab:rq2_variants} presents the resource-efficiency metrics, inference energy,
and pass@1 for all 11 models across the four evaluated variants after five epochs of
self-optimization. Figure~\ref{fig:rq2_energy} summarizes total inference energy per model
and variant. Figure~\ref{fig:efficiency} plots correctness efficiency against normalized
execution time to support the cross-model comparison in RQ2b.

\subsubsection{RQ2a: How do decoding parameters affect energy consumption across models?}
\label{sec:rq2a}

\paragraph{Effect on inference energy.}
As shown in Figure~\ref{fig:rq2_energy}, \textit{mt1024} produces the
lowest total inference energy across all models, with reductions of up to 72\% relative
to \textit{default} (mean 1.54\,MJ vs.\ 2.17\,MJ). The savings are largest for
GLM-4.7-Flash (3.43\,MJ $\to$ 0.95\,MJ, $-$72.2\%) and smallest for Qwen3.5-4B
(0.87\,MJ $\to$ 0.83\,MJ, $-$4.8\%), reflecting how strongly maximum output tokens
constrains generation cost. The \textit{ctx65536} and \textit{t0.5}
variants are broadly comparable to \textit{default} in energy, with \textit{ctx65536}
reaching the highest absolute consumption for Qwen3.5-27B (6.50\,MJ).
\paragraph{Effect on execution quality.}
The \textit{default} variant achieves NET values of 0.95--1.05 across all models,
indicating near-canonical execution time. Non-default variants span a wider range of
0.93--2.51, where the upper extreme is driven by the Nemotron-Nano \textit{mt1024}.
NMU is largely insensitive to variant choice (range 1.002--1.088), while NTMU varies
more substantially, reaching 3.19 for Nemotron-Nano \textit{mt1024} and 1.52 for
Ministral-8B \textit{ctx65536}.

\subsubsection{RQ2b: Which model-configuration pairs offer the most favorable trade-off between inference energy and optimization quality?}
\label{sec:rq2b}

Figure~\ref{fig:efficiency} plots correctness efficiency (pass@1\,/\,$E_{\mathrm{total}}$,
in \%\,MJ$^{-1}$) against NET for all model-variant pairs. The optimal region---low
NET and high correctness per invested joule---is shaded in the upper-left quadrant.

Qwen3.5-4B dominates across all four variants, achieving the highest efficiency of
54--62\,\%\,MJ$^{-1}$ while maintaining NET below 1.0. Its \textit{mt1024} variant
reaches 61.8\,\%\,MJ$^{-1}$, the best single configuration in the study. Ministral-3B
\textit{mt1024} is the best non-Qwen entry in the optimal region at
42.4\,\%\,MJ$^{-1}$. At the opposite extreme, GLM-4.7-Flash and Nemotron-Cascade
\textit{mt1024} are the least efficient configurations (7.3 and 7.5\,\%\,MJ$^{-1}$
respectively), combining high inference cost or correctness collapse with no compensating
gain in execution quality. Qwen3.5-27B illustrates the cost of scaling: despite the
highest absolute pass@1 (80.98\%), its 6.43\,MJ generation cost depresses its efficiency
to $\sim$12.6\,\%\,MJ$^{-1}$, well below Qwen3.5-4B.

\begin{table*}[!ht]
\caption{%
    Resource efficiency, inference energy, and pass@1 after self-optimization.
    N-prefixed metrics are normalized against the canonical solution;
    values below 1.0 indicate better performance than the canonical reference.
    $E_{\mathrm{total}}$ in MJ.
}
\label{tab:rq2_variants}
\centering
\renewcommand{\arraystretch}{0.7}
\setlength{\tabcolsep}{2pt}
\footnotesize
\begin{tabular}{llcccccrrr}
\hline
\textbf{Model} & \textbf{Variant} & \textbf{NET} & \textbf{NMU} & \textbf{NTMU} & \textbf{pass@1 (\%)} & $E_{\mathrm{total}}$ & \textbf{CPU\,\%} & \textbf{DRAM\,\%} & \textbf{GPU\,\%} \\
\hline
  \multirow{4}{*}{GLM-4.6V-Flash} & \textit{default} & 1.038 & 1.003 & 1.040 & 25.4 & 1.050 & 25.6 & 2.5 & 71.9 \\
   & \textit{ctx65536} & 1.031 & 1.002 & 1.033 & 26.1 & 1.034 & 24.7 & 2.2 & 73.1 \\
   & \textit{mt1024} & 1.059 & \textbf{1.002} & 1.058 & 9.0 & \textbf{0.474} & 26.7 & 2.5 & 70.8 \\
   & \textit{t0.5} & 1.028 & 1.003 & 1.028 & 25.4 & 1.050 & 24.3 & 2.4 & 73.3 \\
\hline
  \multirow{4}{*}{GLM-4.7-Flash} & \textit{default} & 0.984 & 1.005 & 0.989 & 25.0 & 3.429 & \textbf{16.1} & 1.6 & 82.3 \\
   & \textit{ctx65536} & 0.966 & 1.004 & 0.971 & 25.5 & 3.442 & 16.7 & \textbf{1.4} & 81.9 \\
   & \textit{mt1024} & 1.007 & 1.004 & 1.009 & 9.3 & 0.954 & 18.4 & 1.8 & 79.8 \\
   & \textit{t0.5} & 0.958 & 1.005 & 0.965 & 27.0 & 3.478 & 16.6 & 1.5 & 81.9 \\
\hline
  \multirow{4}{*}{Ministral-3B} & \textit{default} & 0.946 & 1.009 & 0.956 & 32.7 & 0.871 & 28.5 & 2.8 & 68.7 \\
   & \textit{ctx65536} & 0.952 & 1.009 & 0.961 & 31.6 & 0.844 & 27.0 & 2.5 & 70.4 \\
   & \textit{mt1024} & 0.944 & 1.007 & 0.951 & 31.6 & 0.745 & 29.9 & 2.9 & \textbf{67.2} \\
   & \textit{t0.5} & 0.963 & 1.010 & 0.974 & 30.5 & 0.866 & 28.6 & 2.8 & 68.7 \\
\hline
  \multirow{4}{*}{Ministral-8B} & \textit{default} & 0.949 & 1.015 & 0.966 & 52.6 & 1.635 & 23.9 & 2.1 & 74.0 \\
   & \textit{ctx65536} & 1.338 & 1.033 & 1.524 & 52.3 & 1.637 & 24.1 & 1.8 & 74.1 \\
   & \textit{mt1024} & 0.947 & 1.011 & 0.965 & 52.1 & 1.443 & 24.3 & 2.0 & 73.7 \\
   & \textit{t0.5} & 0.955 & 1.015 & 0.974 & 51.7 & 1.639 & 23.4 & 1.9 & 74.7 \\
\hline
  \multirow{4}{*}{Ministral-14B} & \textit{default} & 0.954 & 1.008 & 0.963 & 55.3 & 2.313 & 23.3 & 1.8 & 74.9 \\
   & \textit{ctx65536} & 0.944 & 1.008 & 0.952 & 56.0 & 2.253 & 20.3 & 1.9 & 77.8 \\
   & \textit{mt1024} & 0.950 & 1.009 & 0.959 & 54.8 & 1.986 & 21.4 & 1.9 & 76.7 \\
   & \textit{t0.5} & \textbf{0.932} & 1.009 & \textbf{0.940} & 54.6 & 2.287 & 21.6 & 1.9 & 76.5 \\
\hline
  \multirow{4}{*}{Nemotron-Cascade} & \textit{default} & 1.050 & 1.019 & 1.079 & 43.2 & 1.651 & 22.0 & 2.0 & 76.0 \\
   & \textit{ctx65536} & 1.001 & 1.014 & 1.017 & 43.7 & 1.667 & 22.8 & 2.0 & 75.2 \\
   & \textit{mt1024} & 1.048 & 1.009 & 1.059 & 4.8 & 0.638 & 21.6 & 2.0 & 76.5 \\
   & \textit{t0.5} & 1.034 & 1.018 & 1.056 & 42.5 & 1.716 & 23.1 & 2.0 & 74.9 \\
\hline
  \multirow{4}{*}{Nemotron-Nano} & \textit{default} & 1.020 & 1.019 & 1.045 & 41.5 & 1.721 & 21.1 & 1.8 & 77.1 \\
   & \textit{ctx65536} & 1.010 & 1.018 & 1.037 & 40.8 & 1.785 & 22.6 & 1.8 & 75.6 \\
   & \textit{mt1024} & 2.507 & 1.088 & 3.189 & 12.5 & 0.784 & 21.6 & 1.7 & 76.7 \\
   & \textit{t0.5} & 1.169 & 1.039 & 1.266 & 40.4 & 1.773 & 23.3 & 1.9 & 74.9 \\
\hline
  \multirow{4}{*}{Qwen3.5-4B} & \textit{default} & 0.972 & 1.010 & 0.979 & 51.3 & 0.871 & 26.4 & 2.5 & 71.1 \\
   & \textit{ctx65536} & 0.973 & 1.011 & 0.979 & 51.4 & 0.858 & 25.3 & 2.3 & 72.4 \\
   & \textit{mt1024} & 0.967 & 1.011 & 0.974 & 51.2 & 0.829 & 27.1 & 2.6 & 70.3 \\
   & \textit{t0.5} & 0.970 & 1.011 & 0.977 & 48.3 & 0.892 & 26.4 & 2.4 & 71.2 \\
\hline
  \multirow{4}{*}{Qwen3.5-9B} & \textit{default} & 1.003 & 1.011 & 1.012 & 66.8 & 1.595 & 22.7 & 2.1 & 75.2 \\
   & \textit{ctx65536} & 0.998 & 1.011 & 1.008 & 66.5 & 1.637 & 22.7 & 2.4 & 74.9 \\
   & \textit{mt1024} & 1.014 & 1.011 & 1.023 & 65.4 & 1.479 & 22.3 & 2.5 & 75.2 \\
   & \textit{t0.5} & 1.000 & 1.011 & 1.008 & 64.5 & 1.618 & 21.7 & 2.1 & 76.1 \\
\hline
  \multirow{4}{*}{Qwen3.5-27B} & \textit{default} & 0.974 & 1.005 & 0.977 & \textbf{81.0} & 6.426 & 18.6 & 1.6 & 79.8 \\
   & \textit{ctx65536} & 0.967 & 1.005 & 0.970 & 80.9 & 6.496 & 19.2 & \textbf{1.4} & 79.4 \\
   & \textit{mt1024} & 0.973 & 1.005 & 0.976 & 79.8 & 5.477 & 18.4 & 1.6 & 80.1 \\
   & \textit{t0.5} & 0.974 & 1.006 & 0.977 & 80.2 & 6.425 & 18.6 & 1.5 & 79.9 \\
\hline
  \multirow{4}{*}{Qwen3.5-35B-A3B} & \textit{default} & 0.971 & 1.006 & 0.974 & 78.6 & 2.314 & 19.5 & 1.7 & 78.8 \\
   & \textit{ctx65536} & 0.977 & 1.006 & 0.980 & 78.6 & 2.377 & 18.7 & 1.9 & 79.4 \\
   & \textit{mt1024} & 0.973 & 1.006 & 0.977 & 77.5 & 2.092 & 23.6 & 2.4 & 74.0 \\
   & \textit{t0.5} & 0.975 & 1.008 & 0.981 & 78.1 & 2.430 & 18.6 & 1.9 & 79.6 \\
\hline
\end{tabular}
\end{table*}

\begin{tcolorbox}[
  colback=gray!8,
  colframe=gray!40,
  boxrule=0.5pt,
  arc=3pt,
  left=6pt, right=6pt, top=2pt, bottom=2pt
]
\textbf{RQ2a:} Constraining output tokens (\textit{mt1024}) is the most effective
knob for reducing inference energy across all models, though it can severely collapse
pass@1 for some configurations.
\textbf{RQ2b:} Qwen3.5-4B dominates the correctness-efficiency frontier across all
variants. Ministral-3B is the strongest non-Qwen alternative. The GLM family is
strictly dominated on both axes, and high-parameter models such as Qwen3.5-27B are
penalized by disproportionate inference cost despite leading on absolute pass@1.
\end{tcolorbox}

\subsection{RQ3: Does the energy invested in generation translate into reduced execution energy when the optimized code is run relative to canonical human reference solutions, and if so, at what number of executions does the investment break even?}
\label{sec:rq3}

\begin{figure*}[!ht]
\centering
\includegraphics[width=0.72\columnwidth]{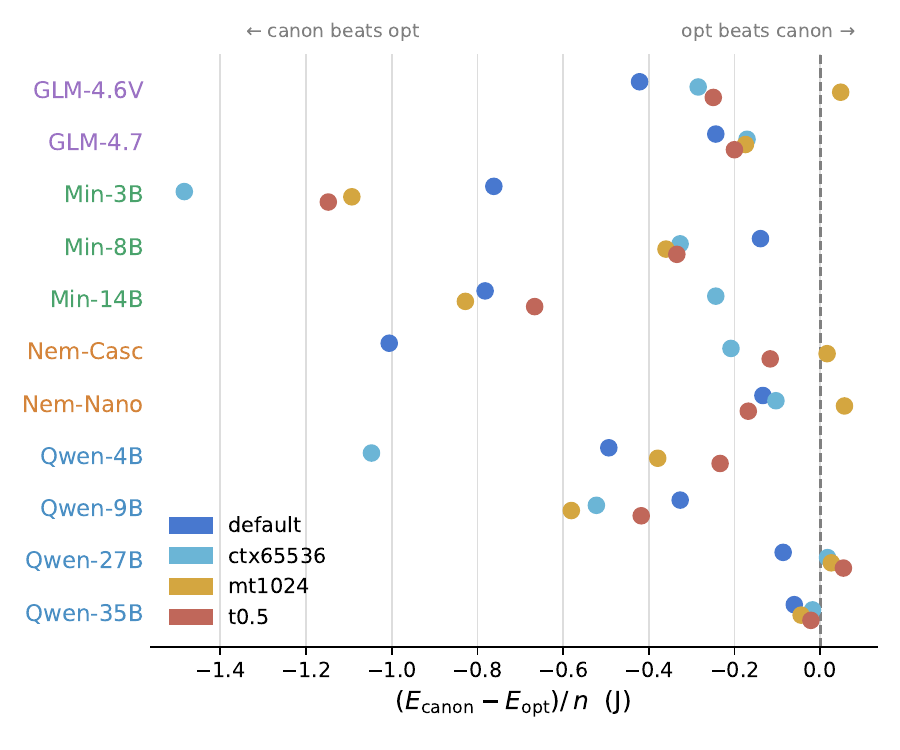}%
\hspace{1em}%
\includegraphics[width=0.72\columnwidth]{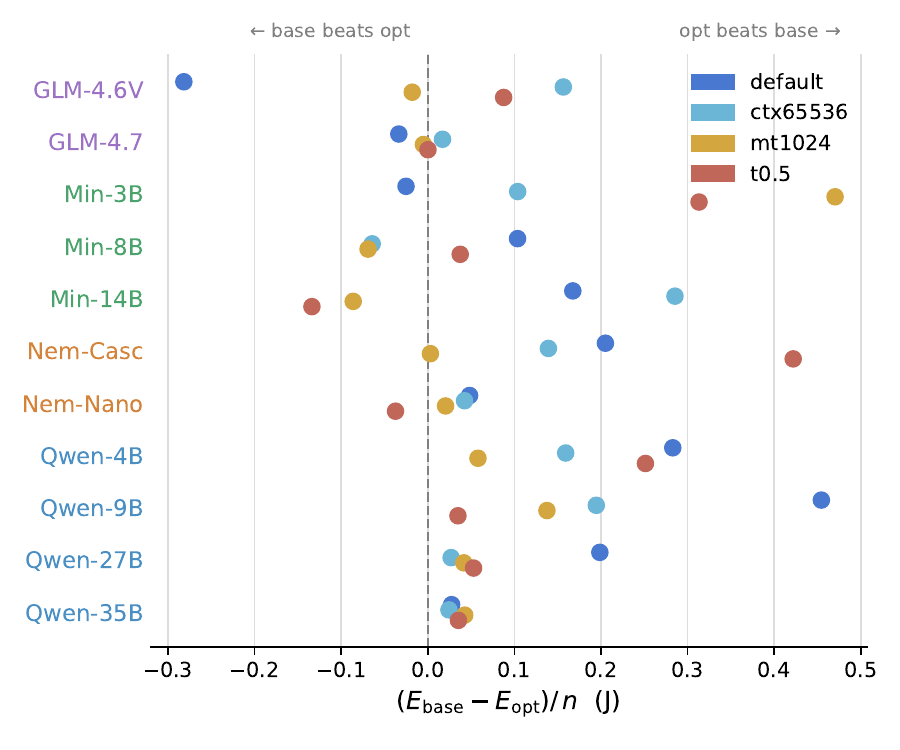}
\caption{%
    \textit{Left:} $(E_{\mathrm{canon}} - E_{\mathrm{opt}})/n$; points right of zero
    beat the canonical baseline.
    \textit{Right:} $(E_{\mathrm{base}} - E_{\mathrm{opt}})/n$; positive values
    indicate improvement over the base epoch.
}
\label{fig:rq3_dots}
\end{figure*}

For each model and decoding
variant we measure the per-execution CPU energy of the last-epoch solution and compare
it against the canonical solution.
Figure~\ref{fig:rq3_dots} (left) reports the per-problem execution energy delta
$(E_{\mathrm{canon}} - E_{\mathrm{opt}})/n$ for all model-variant pairs, and
Figure~\ref{fig:rq3_dots} (right) shows the per-problem spread between the base
and optimized epoch, $(E_{\mathrm{base}} - E_{\mathrm{opt}})/n$.

\paragraph{Canonical solutions dominate at execution time}
Figure~\ref{fig:rq3_dots} (left) shows that 38 of the 44 model-variant configurations
produce optimized code that consumes \emph{more} execution energy per problem than the
canonical reference ($\Delta < 0$). The overheads range from $+6\%$ (Qwen3.5-35B-A3B \textit{ctx65536}) to
$+638\%$ (Ministral-3B \textit{ctx65536}) of the canonical energy. The six exceptions are GLM-4.6V-Flash,
Nemotron-Nano and Nemotron-Cascade under \textit{mt1024}
($0.05$, $0.06$ and $0.02$\,J per problem) and Qwen3.5-27B under \textit{t0.5},
\textit{mt1024} and \textit{ctx65536} ($0.05$, $0.03$ and $0.02$\,J per problem),
the only configurations whose last-epoch solutions
undercut the canonical baseline. The largest overheads are concentrated in the Ministral
family: Ministral-3B across all variants exceeds 0.75\,J per problem, peaking at
1.48\,J (\textit{ctx65536}).
GLM, Nemotron and Qwen variants cluster at the lower end of the distribution
(means 0.21\,J, 0.21\,J and 0.26\,J), while
Ministral sits at the upper end (mean 0.68\,J).

Across decoding configurations, \textit{mt1024} and \textit{t0.5} produce the
lowest mean overheads (0.30\,J and 0.32\,J respectively), compared to \textit{default}
and \textit{ctx65536} (both $\approx$0.40\,J). However, even the most favourable
configuration (Nemotron-Nano \textit{mt1024}) achieves only a marginal gain over the
canonical baseline, and three of the six gains come from \textit{mt1024}
configurations whose pass@1 collapsed below 13\%.

\paragraph{Optimization improves upon its own base generation but does not close the gap}
Figure~\ref{fig:rq3_dots} (right)
reveals that self-optimization reduces execution energy relative to its own
base generation in most cases: 34 of the 44
configurations (77.3\%) show improvement ($\bar{E}_{\text{base}} >
\bar{E}_{\text{opt}}$), while 10 (22.7\%) regress. The Qwen family fares
best at 100\% improved, followed by Nemotron at 87.5\%, while Ministral and GLM
sit at 58.3\% and 50.0\%. The largest
regressions are observed for GLM-4.6V-Flash \textit{default} ($-0.28$\,J per problem)
and Ministral-14B \textit{t0.5} ($-0.13$\,J), where the last epoch produced
more energy-expensive code than the base epoch. However, the mean improvement
of 0.09\,J per problem is well below the mean overhead of 0.36\,J relative to the
canonical solutions, so refinement narrows the gap without closing it.

\paragraph{Break-even analysis}
A break-even point $N^{*}$ exists only when the optimized code consumes less
execution energy than the canonical solution on average. Under this criterion,
38 of the 44 configurations admit \emph{no break-even point}: the energy
invested in generation is never recovered regardless of how often the resulting
code is executed. For the six configurations where a saving exists, the
thresholds remain large. Their break-even point has a median of approximately
$2.1 \times 10^{5}$ executions per problem (range $1.0 \times 10^{5}$ to
$8.4 \times 10^{5}$), thresholds reachable only in high-volume deployment
scenarios. We
discuss the broader implications for the energy economics of LLM-based code
optimization in Section~\ref{sec:discussion}.

\begin{tcolorbox}[
  breakable,
  colback=gray!8,
  colframe=gray!40,
  boxrule=0.5pt,
  arc=3pt,
  left=6pt, right=6pt, top=5pt, bottom=5pt
]
\textbf{RQ3:}
Self-optimized code almost universally consumes more execution energy than the
human canonical reference, with only six of 44 configurations achieving a marginal gain.
Most configurations improve relative to their own base epoch, but not enough to
reach the canonical reference. Where a break-even
point exists, it requires hundreds of thousands of executions, making LLM-based self-optimization
a net energy loss in all but high-volume deployment scenarios.
\end{tcolorbox}

\subsection{Carbon Footprint of the Optimization Loop}
\label{sec:carbon}

To translate energy measurements into sustainability-relevant units, we convert
inference energy into operational carbon emissions using the global average grid
intensity of $\gamma = 217$\,gCO\textsubscript{2}eq/kWh from Cheung et
al.~\cite{cheung2025comparative} ($\gamma = 60.3\,\mu$gCO\textsubscript{2}eq/J).
This value lies within the 23.92--625.57\,kg/MWh range across 365 regional grids
tracked by eco2AI~\cite{budennyy2022eco2ai}. Region-specific estimates can be
obtained by scaling our figures by $\gamma_{\text{local}}/217$.

\paragraph{Generation-side footprint}
Table~\ref{tab:carbon_gen} reports emissions for the default configuration.
Total footprint ranges from 52.5\,gCO\textsubscript{2}eq (Ministral-3B and
Qwen3.5-4B) to 387.3\,gCO\textsubscript{2}eq (Qwen3.5-27B) for optimizing
1{,}000 problems.

\begin{table}[!ht]
\caption{%
    Carbon footprint of the five-epoch optimization loop (default variant), computed using
    $\gamma = 217$\,gCO\textsubscript{2}eq/kWh~\cite{cheung2025comparative}.
}
\label{tab:carbon_gen}
\centering
\setlength{\tabcolsep}{4pt}
\begin{tabular}{lrr}
\hline
\textbf{Model}
  & $E_{\mathrm{total}}$ \textbf{(MJ)}
  & \textbf{gCO\textsubscript{2}eq} \\
\hline
GLM-4.6V-Flash   & 1.050 &  63.3 \\
GLM-4.7-Flash    & 3.429 & 206.7 \\
\textbf{Ministral-3B}     & \textbf{0.871} & \textbf{52.5} \\
Ministral-8B     & 1.635 &  98.5 \\
Ministral-14B    & 2.313 & 139.4 \\
Nemotron-Cascade & 1.651 &  99.5 \\
Nemotron-Nano    & 1.721 & 103.7 \\
\textbf{Qwen3.5-4B}       & \textbf{0.871} & \textbf{52.5} \\
Qwen3.5-9B       & 1.595 &  96.1 \\
Qwen3.5-27B      & 6.426 & 387.3 \\
Qwen3.5-35B-A3B  & 2.314 & 139.5 \\
\hline
\end{tabular}
\end{table}

\paragraph{Execution-side footprint and break-even in carbon units}
Execution-side emissions are roughly three orders of magnitude smaller than generation.
For the six configurations where the optimized solution undercuts the canonical
baseline (Section~\ref{sec:rq3}), the per-execution savings in carbon terms are
negligible: GLM-4.6V-Flash \textit{mt1024}, the fastest to break even, saves
$\sim$2.9\,$\mu$gCO\textsubscript{2}eq
per problem per execution, against a generation investment of
28.6\,gCO\textsubscript{2}eq, yielding $N^{*} \approx 1.0 \times 10^{5}$
executions, which is identical to the energy-denominated result since the conversion is
linear in $\gamma$.

\section{Discussion}
\label{sec:discussion}

We now interpret the results across the three research questions and their
broader implications for practitioners considering LLM-based code optimization.

\subsection{Generation energy as a hidden cost of optimization}

A central finding of this study is that the energy consumed during LLM-based
self-optimization is substantial and largely unrecoverable. Across 38 of the 44
evaluated model-variant configurations, the optimized code consumes more
execution energy per problem than the canonical human-written reference, and the
generation investment requires hundreds of thousands of executions to break even in the six
cases where a saving exists. Our results suggest that performance optimization
and energy efficiency are not aligned goals: faster code does not automatically
mean greener code when the cost of generation is accounted for. Coignion et
al.~\cite{coignion2025faster} reached a similar conclusion on a different
benchmark and pipeline, reporting weak correlation between performance gains and
energy reduction.

\subsection{Model and architecture choice matter more than decoding parameters}

Our RQ2 analysis shows that the dominant factor in both inference energy and
correctness-efficiency is the choice of model and its architecture, not decoding configuration.
Constraining output length (\textit{mt1024}) consistently reduces inference
energy across all models, but at the cost of correctness collapse in several
configurations. The MoE architecture of Qwen3.5-35B-A3B achieves near-identical
correctness to the Qwen3.5-27B dense model at less than half the
inference energy, owing to sparse parameter activation, suggesting that
architectural choices made upstream of deployment have a larger sustainability
impact than parameter tuning.

\subsection{Self-optimization reduces execution energy across epochs, but not to the canonical level}

A subtler finding from RQ3 is that the EffiLearner pipeline does drive
execution-energy reductions relative to its own base epoch in 77.3\% of
configurations, yet 22.7\% still regress across the optimization trajectory, and
the mean improvement of 0.09\,J per problem recovers only a fraction of the mean
0.36\,J gap to the canonical solutions. Future
pipelines should incorporate execution energy as an explicit selection criterion
alongside correctness.

\subsection{Practical implications}

LLM-based self-optimization is only energetically justified for code that will
be executed at very high volume. For one-off scripts or lightly used utilities,
the generation cost will never be recouped. Practitioners aiming to reduce overall energy consumption should restrict
optimization to hot-path code with known high execution frequency and prefer
smaller models such as Ministral-3B or Qwen3.5-4B when correctness requirements
allow. The carbon footprint analysis (Section~\ref{sec:carbon}) confirms that
these differences are non-trivial at scale: the generation footprint of
Qwen3.5-27B is approximately seven times that of Ministral-3B for the same
optimization task.
\section{Limitations and Future Work}
\label{sec:limitations}
The findings reported in this study are bounded by deliberate experimental
choices, each pointing to a concrete direction for future work.

\paragraph{Benchmark scope}
EffiBench problems are short, algorithmically dense, and complete in under a
second, which is ideal for energy measurement at scale but unrepresentative of
long-running workloads where break-even would most plausibly favor optimization.
Future work should investigate whether the pass@1 collapse observed under
aggressive decoding constraints persists on harder, more diverse tasks such as
I/O-bound programs or data-processing pipelines.

\paragraph{Human baseline}
The canonical EffiBench solutions are top-rated submissions from a competitive
programming forum, so they represent highly efficient human implementations
rather than the code an average practitioner would write. Our RQ3 comparison
therefore measures LLM-optimized code against a strong reference. We also assign no
authoring cost to the canonical solutions, whereas the generation cost of the
LLM is fully counted. This omission does not affect our main finding. In 38 of
44 configurations the optimized code consumes more execution energy per run
than the canonical code, so the recurring difference is negative and no
authoring cost for the baseline can turn the generation energy into a net
saving through repeated execution. It could only favor LLM-based optimization
when a solution is executed few times. Future work should compare against
solutions from a broader pool of practitioners and measure human authoring
effort explicitly.

\paragraph{Pipeline design}
The EffiLearner pipeline replaces solutions with no gate on execution energy,
which allows the regressions we observed in 22.7\% of configurations and makes
it difficult to attribute results solely to model capability. Future pipelines
should incorporate execution energy as an explicit selection criterion to isolate
its effect.

\paragraph{Language and platform coverage}
All experiments use Python, whose interpreted execution model shapes the energy
profiles we report. The execution-energy gap between LLM-generated and
canonical solutions may narrow in compiled languages such as C\texttt{++} or
Rust~\cite{pereira2017energy}, and a multi-language replication would establish
whether our patterns generalize.

\paragraph{Closed-source models}
Our open-source-only model selection is required by the need for hardware-level
energy attribution, so our findings do not extend to frontier models served
through commercial APIs. Closing this gap will require provider-published
per-request energy disclosures or instrumented inference stacks.

\paragraph{Broader environmental impact}
Our measurements cover operational energy only. Embodied costs of GPU
manufacturing, data-center cooling, model training, and hardware refresh cycles
are not captured yet can outweigh operational savings over the hardware
lifecycle. Future work should adopt a lifecycle perspective accounting for these
indirect impacts.
\section{Conclusion}
\label{sec:conclusion}
This paper presented a systematic measurement study of the energy economics of
LLM-based code self-optimization, covering inference and execution cost across
11 open-source models, four decoding configurations, and 1{,}000 EffiBench
problems. The central question was whether performance-optimized code produced
by LLM self-refinement is also energy-efficient relative to canonical solutions.

Optimized code consumes more execution energy than the canonical reference in
38 of 44 model-variant pairs; the six exceptions achieve only marginal gains,
with a median break-even point of about $2 \times 10^{5}$ executions per
problem. Refinement does help within its own
trajectory: 77.3\% of configurations reduce execution energy
relative to their base epoch, but the average improvement closes only a
fraction of the gap to the canonical solutions, and the remaining regressions
reflect the absence of any energy-aware
selection criterion in the pipeline.

On the generation side, inference energy varies by nearly an order of magnitude
across models, driven primarily by model size and architecture. MoE models such
as Qwen3.5-35B-A3B offer substantially lower inference cost than
equivalently-sized dense counterparts without sacrificing correctness, making
architectural choice a more consequential sustainability lever than decoding
parameters. Constraining output tokens (\textit{mt1024}) is the most effective
parameter adjustment for reducing inference energy, but risks correctness
collapse that must be weighed per model.

Practitioners should restrict LLM-based self-optimization to hot-path code with
reliably high execution frequency, prefer small and MoE-based models when
correctness allows, and treat generation energy as a first-class cost. Future
work should extend this framework to compiled languages, long-running workloads,
and energy-aware refinement pipelines.

\section*{Acknowledgment}
\textbf{Funding.}
This research was supported by ITEA GreenCode, ITEA Genius and the Federal Ministry of
Research, Technology and Space of Germany under grant number 16IS24070G.
\textbf{Transparency note.}
The readability of this text has been improved with the support of AI
technologies, which were also used in the code generation process.
After using these tools, the authors reviewed and edited the content as needed
and take full responsibility for the content of the publication.

\bibliographystyle{IEEEtran}
\bibliography{references}
\end{document}